\documentclass[sigconf,nonacm]{acmart}

\usepackage{amsmath}
\usepackage{booktabs}
\usepackage{multirow}
\usepackage{xspace}
\usepackage{makecell}
\usepackage{xcolor}
\usepackage{graphicx}
\usepackage{eso-pic}

\newcommand{\diff}[1]{\,{\scriptsize\textcolor{gray}{$(#1)$}}}

\AtBeginDocument{}

 \newcommand{\PaperTitle}{%
    PailitaoMMSearch2.0 Technical Report: Scaling Generative Retrieval to Unified End-to-End
    Multimodal Search%
  }

  \newlength{\AlibabaLogoLift}
\begin{document}


  \AddToShipoutPictureFG*{%
    \AtTextUpperLeft{%
      \put(0,\LenToUnit{\AlibabaLogoLift}){%
        \makebox(0,0)[lt]{%
          \begin{minipage}{\textwidth}
            \raggedright
            \includegraphics[width=3.0cm]{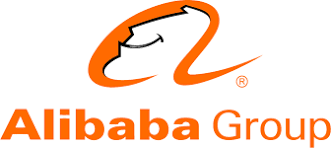}\par
            \vspace{-3pt}
            \noindent\rule{\textwidth}{0.6pt}
          \end{minipage}%
        }%
      }%
    }%
  }

  \title[\PaperTitle]{\PaperTitle}



 \author{%
      \begin{tabular}{c}
        Xiaohan Ye, Zihan Gong, Jingqian Zhao, Lianyu Du, Yunmeng Shu, Xu Chen\textsuperscript{*}, \\
        Zesheng Zhou, Renchun You, Bocheng Chen, Jinsong Lan, Haihong Tang, Bo Zheng
      \end{tabular}%
  }

  \affiliation{%
      \institution{Taobao \& Tmall Group of Alibaba}
      \city{}
      \country{}
  }

  \email{\{yxh268746,huaisong.cx\}@alibaba-inc.com}

  \renewcommand{\shortauthors}{Ye et al.}

\begin{abstract}
Generative retrieval models have been increasingly adopted in search and recommendation systems. 
However, industrial generative retrieval models are commonly specialized to a fixed online scenario: they learn from scenario-specific interactions, support a narrow set of query forms and retrieval objectives. We study how to remove these couplings and build a unified end-to-end
multimodal search system for open-ended image--text queries, heterogeneous
shopping intents, and online scenarios with different resource constraints. We
present \textbf{PailitaoMMSearch}, which directly generates product Semantic
IDs from a raw query image and free-form instruction and scales along three
dimensions. \textbf{Data scaling} uses a reusable toolset that can be composed
into intent- and lifecycle-specific workflows for cold-start synthesis, quality
control, real-traffic refinement, and continuous data production.
\textbf{Intent scaling} progressively equips one model with e-commerce product
knowledge, multiple retrieval relations, and adjustable reasoning budgets through
continual pre-training (CPT), mixed-budget supervised fine-tuning (SFT), and
unified reinforcement learning (RL). The RL stage combines intent-balanced
optimization and Sample--Beam rollouts
aligned with multi-candidate serving. \textbf{Scenario scaling} exposes
quality--cost operating points across model capacity and reasoning budget. The
system is deployed in multimodal search applications on Taobao Pailitao. A
controlled online A/B test reports statistically significant improvements,
including item page views $\textbf{+17.90\%}$, transactions $\textbf{+27.04\%}$, and gross
merchandise value $\textbf{+8.94\%}$.
\end{abstract}



\maketitle

\begingroup
  \renewcommand{\thefootnote}{\fnsymbol{footnote}}
  \footnotetext[1]{Corresponding Author and Project Leader.}
  \endgroup

\section{Introduction}
\label{sec:intro}

Generative retrieval represents catalog items as token sequences and retrieves
an item by autoregressively generating its identifier. This formulation turns
retrieval into sequence modeling, enables items to share token-level semantics,
and avoids scoring every query against the full catalog
\citep{dsi,genre,tiger}. Recent industrial systems further show that Semantic
ID (SID) generation can serve large-scale search and recommendation traffic
\citep{chen2025onesearch,zhou2025onerecv2technicalreport}. Despite this
progress, most deployed generative search models remain closely coupled to the
online scenario for which they were built.

The coupling appears in three forms. First, \emph{supervision binding} occurs
when training is dominated by interactions from an existing entry point. Such
logs reflect the query format, exposed products, and intent distribution of
that entry point, they provide little supervision for an intent that has not yet
been launched. Second, \emph{intent binding} occurs when one model is trained to
fit one retrieval relation, such as visual similarity or recommendation.
Supporting another relation then requires an intent classifier, a routing rule,
or another expert model, which inflates system complexity and maintenance cost
and lets routing errors propagate into the downstream retrieval result. Third,
\emph{resource binding} occurs when model size, decoding strategy, and inference
budget are selected for one service-level objective. The resulting configuration
is difficult to reuse across a high-throughput candidate-generation service, a
latency-tolerant complex search application, and the many operating points
between them.

Open-ended multimodal product search exposes all three limitations at once, and
in their most severe form. A
user may upload a shirt and ask for ``sports shorts to match it,'' photograph a
pair of sandals and request ``the women's version from this brand,'' or capture
a household product and ask for ``a foldable one for children.'' These queries
share the same image--text interface but express different retrieval relations.
The desired result may preserve the pictured product, modify selected
attributes, or belong to a complementary category. Query complexity also varies
within an application: some requests can be answered from direct visual and
attribute evidence, while others require multi-step comparison of category,
style, function, and constraints. A practical system should understand these
relations without requiring the user or an upstream router to name the intent.

We formulate the resulting problem as \emph{unified open-ended multimodal
generative retrieval}: one model consumes a raw image and a free-form textual
instruction, infers their implicit retrieval relation, and generates a set of
product SIDs. We instantiate this formulation in
\textbf{PailitaoMMSearch}, a native vision-language model (VLM) system that
scales along three connected dimensions:

\begin{itemize}
    \item \textbf{Data scaling broadens supervision coverage.} A reusable
    toolset performs candidate screening, relation verification, query
    generation or repair, reasoning annotation, and terminal quality control.
    The tools are composed into different workflows according to the target
    intent, source data, and lifecycle stage, from cold start to real-traffic
    refinement and continuous production.

    \item \textbf{Intent scaling converts heterogeneous supervision into a
    unified retrieval policy.} Continual pre-training (CPT) first adapts the VLM to e-commerce products
    and the SID output space. Multi-intent, mixed-budget supervised fine-tuning (SFT) then establishes
    similar, editing, and matching behaviors together with several reasoning
    budgets. Unified reinforcement learning (RL) then optimizes all intents jointly
with Sample--Beam rollouts, which sample the reasoning trajectory but
beam-decode the product SIDs, so that policy optimization matches the
candidate-set generation used online.

    \item \textbf{Scenario scaling converts unified capability into deployable
    operating points.} Model capacity provides coarse-grained adaptation to an
    application's resource envelope, while reasoning budget provides
    fine-grained adaptation to query complexity and latency requirements. These
    operating points expose the behaviors already learned during intent scaling,
    so no separate routing model is introduced.
\end{itemize}

We evaluate the system through five questions: overall multi-intent retrieval
quality; the effects of data volume, quality control, and real-traffic
onboarding; the contributions of CPT, multi-intent transfer, reasoning-budget
SFT, and RL; the quality--cost space induced by model and reasoning scale; and
online business impact. The system has been deployed in multimodal search
applications on Taobao Pailitao. Our contributions are:

\begin{enumerate}
    \item We formulate the transition from scenario-specific generative search
    to unified open-ended multimodal retrieval and implement a native-VLM system
    that maps free-form image--text queries directly to product SIDs without
    explicit intent input or intent-specific serving routes. More broadly, we
    provide a practical, industrial-scale blueprint for building open-ended
    multimodal search systems, spanning multimodal training-data synthesis and
    quality control, multi-stage model training, and production serving.

    \item We develop a three-dimensional scaling framework: a composable data
    toolset supports intent- and lifecycle-specific workflows; a multi-stage training recipe aligns heterogeneous retrieval relations in one model; and model capacity
    together with reasoning budget forms deployable quality--cost operating
    points.

    \item We introduce Sample--Beam RL rollouts and validate the complete system
    with offline and human evaluation, controlled scaling studies, and
    production measurements. Sample--Beam RL raises the macro Excellent Rate by $5.97$ points
    over SFT, and a controlled online A/B test on Taobao Pailitao yields
    statistically significant gains, including $+17.90\%$ item page views,
    $+27.04\%$ transactions, and $+8.94\%$ gross merchandise value.
\end{enumerate}

\section{Related Work}
\label{sec:related}

\textbf{Generative retrieval and Semantic IDs.}
Generative retrieval treats a model's parameters as a differentiable index and
decodes document, entity, or item identifiers conditioned on a query
\citep{dsi,genre,tiger}. Residual quantization provides hierarchical item tokens
that share coarse-to-fine content semantics \citep{rqvae}, while later work
integrates collaborative signals and richer identifier construction
\citep{lcrec,fu2026forge}. Unified recommendation and industrial systems have
extended the formulation across tasks and production traffic
\citep{p5,chen2025onesearch,zhou2025onerecv2technicalreport}. Our focus differs
from improving the identifier alone: we study how a SID-based retriever can be
scaled beyond one fixed query distribution, retrieval relation, and serving
configuration.

\textbf{Multimodal product retrieval.}
CLIP-style contrastive learning has enabled large-scale visual and cross-modal
product retrieval \citep{clip,ebay_visual_search,pinterest_visual_search}, and
native VLMs allow pixels and language tokens to interact throughout the model
\citep{liu2024llava,internvl,qwen2vl}. Product-specific representation models
further adapt multimodal encoders to catalog semantics
\citep{itemsage,eproduct,productgpt}. These methods commonly retrieve by vector
similarity or specialize in a predefined composed-retrieval task. We instead
use a native VLM to generate catalog SIDs and train one policy over overlapping
retrieval relations expressed through unrestricted image--text instructions.

\textbf{Model adaptation and adaptive reasoning.}
CPT adapts a foundation model to a target domain, while instruction tuning and
RL align model behavior with downstream objectives
\citep{instructgpt,dpo,grpo}. Chain-of-thought methods improve difficult
reasoning tasks, and inference-time methods trade additional computation for
quality \citep{wei2022chain,selfconsistency,tot,s1}. Our mixed-budget SFT establishes
multiple search behaviors under explicit, prompt-controlled reasoning budgets.
Unlike general text reasoning, the action ultimately represents a beam of
catalog identifiers, motivating a rollout that separates sampled reasoning from
beam SID generation.

\begin{figure*}[t]
    \centering
    \includegraphics[width=\textwidth]{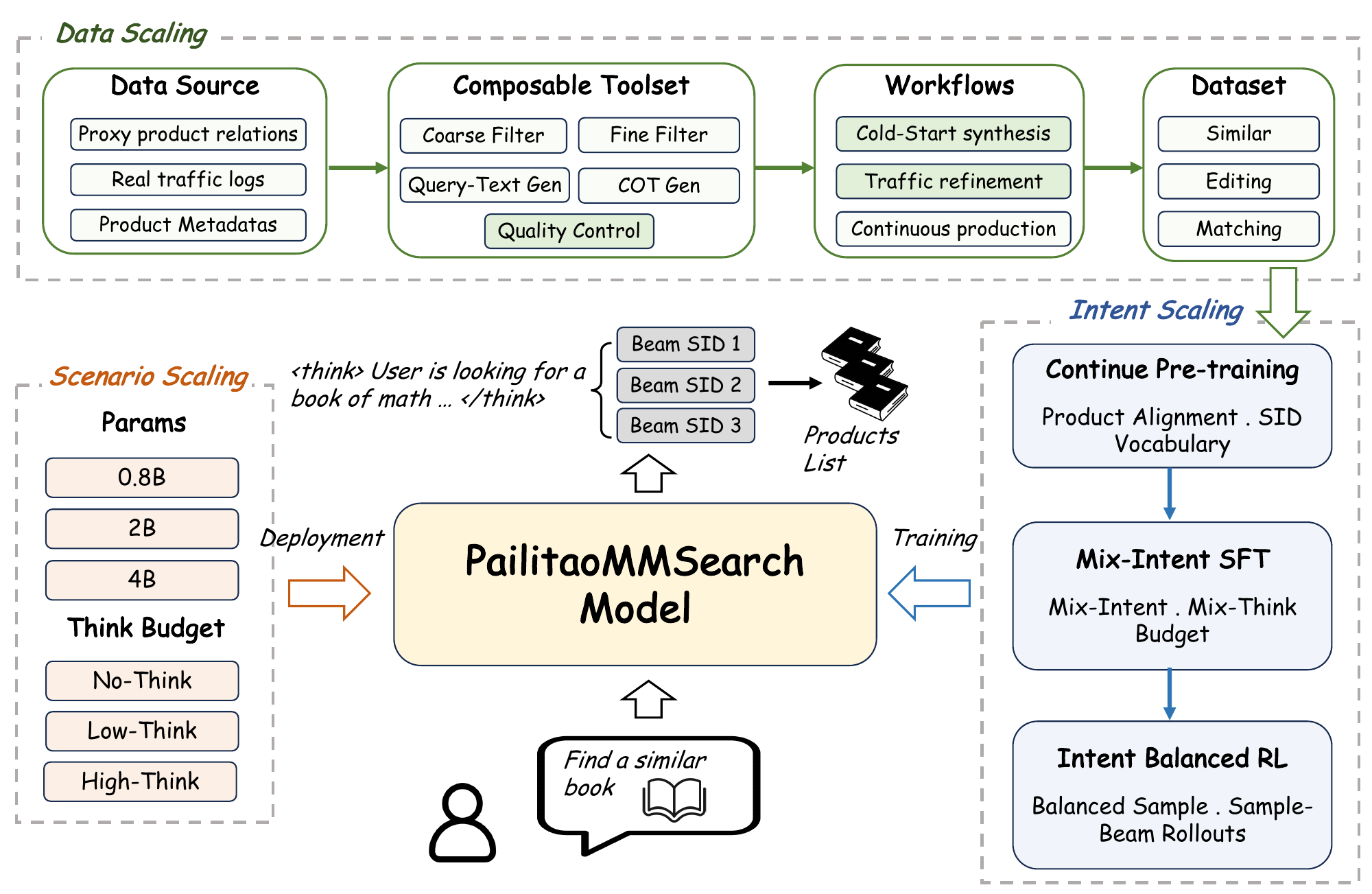}
    \vspace{-10pt}
    \caption{PailitaoMMSearch2.0 overview. Data scaling expands supervision,
    intent scaling learns a unified policy, and scenario scaling exposes the
    learned capability under different resource envelopes.}
    \label{fig:overview}
\vspace{-8pt}
\end{figure*}

\section{Problem and System Overview}
\label{sec:overview}

\subsection{Unified Multimodal Generative Retrieval}

Let a request be
\begin{equation}
    x=(I,Q,C),
\end{equation}
where $I$ is a query image, $Q$ is a free-form textual instruction, and $C$
denotes optional application context. Each catalog item $v$ is assigned a
three-level hierarchical SID $s(v)=(s_1,s_2,s_3)$, with a codebook of $8{,}192$
entries at each level. The model $\pi_\theta$ generates a candidate set
\begin{equation}
    \mathcal{Y}=\operatorname{Decode}_B(\pi_\theta(\cdot\mid x)),
\end{equation}
where $B$ is the serving beam width and every valid sequence maps
deterministically to a catalog item. Decoding is therefore the retrieval
operation rather than a query-rewriting stage for a separate retriever.

The request has an unobserved retrieval relation $z$ and an appropriate
reasoning budget $b$. In our production data, the main evaluation slices are
\emph{similar}, \emph{editing}, and \emph{matching}. Similar queries preserve
most visual and semantic properties, editing queries keep the product identity
while changing or filtering selected attributes, matching queries seek a
complementary item. These slices organize data and evaluation, but they are not
required inputs to the serving model and need not be mutually exclusive. The
system must infer the retrieval relation $z$ from $x$.

For deployment scenario $a$, let $c=(m,b,B)$ denote a serving configuration
consisting of model scale, reasoning budget, and beam width. The system seeks a
shared policy and a scenario configuration that maximize retrieval utility
under resource constraints:
\begin{equation}
\begin{aligned}
(\theta^*,c_a^*)
&= \arg\max_{\theta,c}\;
   \mathbb{E}_{x\sim\mathcal{D}_a}
   \bigl[r_z(x,\mathcal{Y}_{\theta,c})\bigr] \\
\text{s.t.}\quad
& L_{95}(c)\leq\tau_a,\qquad M(c)\leq\mu_a.
\end{aligned}
\label{eq:system_objective}
\end{equation}
where $r_z$ is the relevance rubric implied by the query relation, $L_{95}$ is
P95 latency, $M$ is serving memory, and $(\tau_a,\mu_a)$ is the application's
resource envelope. This expression is a system-level formulation rather than a
single differentiable training loss. Data scaling changes the training
distribution available for estimating $r_z$, intent scaling learns the shared
$\theta$, scenario scaling selects $c_a$ without changing the query interface or
introducing an intent-specific expert.

\vspace{-4pt}
\subsection{Overall Architecture}

PailitaoMMSearch2.0 uses a native VLM from the Qwen3.5
\citep{qwen3.5}. Raw image patches and text tokens are jointly
processed before the model emits an optional \texttt{think} block and one or
more product SIDs. Compared with feeding a fixed image embedding into a language
model, native multimodal attention preserves fine-grained evidence needed to
resolve text-conditioned attributes and relations. All intents share the VLM,
SID vocabulary, and decoding interface.


Figure~\ref{fig:overview} summarizes the dependency among the three scaling
dimensions. Data scaling supplies broad, quality-controlled supervision. Intent
scaling turns the supervision into one model that understands heterogeneous
retrieval relations and selects its reasoning behavior. Scenario scaling then
measures and deploys the resulting policy at different capacity and reasoning
budgets. This order separates how a behavior is learned from how it is used
under production constraints.

\vspace{-10pt}
\section{Data Scaling with Composable Toolsets}
\label{sec:data}

\subsection{Data Sources and Lifecycle Challenges}

Open-ended image–text search cannot rely solely on mature interaction logs. Before a new query format or intent is launched, no direct traffic or interaction data are available, making it necessary to draw on proxy sources. These sources provide useful product-to-product relations, but vary in both structure and reliability. Such relations may be derived from product similarity, co-occurrence patterns, outfit compatibility, and other signals, while the underlying product data may include images, titles, categories, attributes, and other modalities. Once the application is launched and begins receiving traffic, interaction logs offer additional information. However, early user interactions are typically sparse and noisy, often containing misspellings, incomplete constraints, accidental clicks, and exploratory behavior.

We therefore define data scaling over volume, query and intent coverage,
lifecycle coverage, and quality. The goal is not one fixed data pipeline. It is
to reuse a set of operations across diverse intents and stages, so a new
workflow can be assembled without rebuilding the entire production process.

\vspace{-8pt}
\subsection{Reusable Toolset}

The toolset contains five operations with different cost and precision profiles.
They form a configurable processing graph: a workflow may invoke, skip, or
revisit a tool according to its source and target supervision.

\begin{itemize}
    \item \textsc{Coarse-Filter} cheaply removes implausible query--item or
    item--item pairs using category, attribute, behavioral, and lightweight
    multimodal signals. Its thresholds and retained distribution are
    intent-conditioned.

    \item \textsc{Fine-Filter} verifies whether a retained pair expresses the
    target relation. It checks, for example, whether an editing pair preserves
    product identity while changing the requested attribute, or whether a
    matching pair is complementary rather than merely similar.

    \item \textsc{Query-Text-Gen} converts proxy relations into free-form
    image--text requests. For real traffic, the same interface performs minimal
    repair of spelling, incomplete expressions, and conflicting constraints
    while preserving the inferred user intent.

    \item \textsc{CoT-Gen} generates concise or detailed chain-of-thought (CoT) supervision
    when required by the SFT configuration. It grounds the target relation in
    visible evidence, text constraints, and product knowledge rather than
    merely explaining the selected item after the fact.

    \item \textsc{Quality-Control} is the terminal gate. A multimodal evaluator
    applies an intent-specific rubric to the complete query--target instance.
    Its acceptance thresholds are calibrated against held-out human labels, and
    rejected or uncertain examples may be sent back to an earlier tool.
\end{itemize}

The reusable interface is important because the same operation has different
semantics across intents. Similarity filtering emphasizes preserved visual and
category evidence, editing checks invariant and changed attributes, matching
checks compatibility, function, and context. Adding an intent therefore
requires a relation specification and quality rubric plus a workflow
configuration, rather than a separate end-to-end annotation system.

\paragraph{Cost-aware orchestration.}
The tools are arranged as a cascade rather than applying the strongest
multimodal model to every raw pair. Cheap structured and textual checks first
reduce a large candidate pool, fine-grained visual-language verification and
query generation are invoked only on retained examples. The orchestrator also
uses the presence of source fields to determine the path. A product pair with
no user text requires query synthesis, whereas a traffic record normally keeps
the original instruction and invokes repair only when a quality check detects a
recoverable defect. CoT generation is bypassed for No-think supervision and
called with the requested budget for reasoning examples. This conditional
execution reduces production cost and prevents unnecessary rewriting of real
queries.

\paragraph{Quality calibration and production control.}
Every completed instance passes the \textsc{Quality-Control} module. We manually annotated a calibration dataset independent of the training and test sets to adjust the rubric of the module. By scoring the calibration dataset, comparing the consistency with human annotations, and analyzing the patterns of scoring errors across different intents, the rubric settings were updated, enabling this module to serve as a reasonably effective substitute for human annotators in achieving rapid quality control aligned with human preferences. We trace how this rubric evolved for the
editing intent in Appendix~\ref{app:rubric-evolution}.

\subsection{Intent- and Lifecycle-Specific Workflows}

We compose the tools into three recurring workflows.
\textbf{Cold-start synthesis} begins with proxy product relations, applies
coarse and fine verification, generates an image--text request, optionally
adds reasoning, and passes the instance through quality control. It creates
training data before direct traffic is available. \textbf{Real-traffic
refinement} starts from observed query--item interactions, removes unreliable
behavior, repairs the query when necessary, optionally annotates reasoning, and
applies the same terminal gate. It preserves authentic language and demand
while reducing noise. \textbf{Continuous production} mixes proxy sources with
accumulated traffic, periodically samples hard or under-covered regions, and
reuses the tools for new catalog items, expressions, and intents.

Table~\ref{tab:workflows} summarizes the composition. A checkmark denotes a
standard operation and a circle denotes a conditional operation selected by the
source quality or training configuration. Quality-Control remains mandatory in
all cases, which makes the accepted corpora comparable even though their source
paths differ.

\begin{table}[t]
\centering
\caption{Tool composition for the three data-production workflows. QTG and QC
denote Query-Text-Gen and Quality-Control.}
\vspace{-10pt}
\label{tab:workflows}
\small
\begin{tabular}{lccccc}
\toprule
Workflow & Coarse & Fine & QTG & CoT & QC\\
\midrule
Cold-start synthesis & \checkmark & \checkmark & \checkmark & $\circ$ & \checkmark\\
Real-traffic refinement & $\circ$ & \checkmark & $\circ$ & $\circ$ & \checkmark\\
Continuous production & \checkmark & \checkmark & $\circ$ & $\circ$ & \checkmark\\
\bottomrule
\end{tabular}
\vspace{-12pt}
\end{table}

\paragraph{Worked editing example.}
Consider a proxy pair consisting of a red short-sleeve shirt as the query
product and a red long-sleeve shirt as the target. Coarse-Filter retains the
pair because category and dominant appearance are compatible. Fine-Filter then
produces a structured relation: category and color are invariants, sleeve
length is the intended edit, and other uncontrolled differences must remain
below the acceptance threshold. Query-Text-Gen verbalizes only the supported
transformation, such as ``find a long-sleeve one in the same color,'' rather
than describing every target attribute. For a Low-think example, CoT-Gen
records the compact decision that the query preserves category and color while
changing sleeve length, for No-think data it is skipped. Quality-Control
finally evaluates the source image, generated instruction, and target together.
A pair is rejected if the target changes category, violates the preserved
color, or contains additional differences that make the instruction
underspecified. The same tools produce matching data with a different relation
rubric: category change may be required, while compatibility and context replace
attribute preservation as the key acceptance conditions.

This composition supports both axes required by industrial expansion: diverse
retrieval relations and different points in an application's lifecycle. It
also separates source expansion from quality guarantees. Cold-start data enables
launch, launch produces real queries, and refined traffic improves subsequent
model versions without discarding the proxy data that covers rare relations.

\vspace{-8pt}
\section{Intent Scaling via Multi-Stage Training}
\label{sec:intent}

\subsection{Continual Pre-Training}

Generic vision–language models (VLMs) possess limited prior knowledge of platform-specific product taxonomies, attribute schemas, and product SID vocabularies. To bridge this gap, we perform continued pre-training (CPT) on a mixture of e-commerce-specific and general-domain corpora. The training tasks encompass product-title understanding, product SID prediction, product-attribute prediction, e-commerce user-behavior prediction, and general natural-language modeling. This stage enables the model to acquire a unified understanding of e-commerce products across visual and textual modalities, while providing a stronger initialization for the SID vocabulary space used in generative retrieval.

\subsection{Multi-Intent, Mixed-Budget SFT}
Through the data-construction workflow, we obtain multi-intent training data. Each training sample maps an open-ended query to the SID of a target product and may additionally include a chain-of-thought (CoT) rationale.

The training samples are organized into similarity, editing, and matching subsets. Nevertheless, all samples jointly update a single model and share a unified input–output interface. At serving time, the model is not provided with explicit intent labels. Instead, it must infer the appropriate retrieval relation from the evidence conveyed by the instruction and the accompanying images.

Based on the form and granularity of the CoT rationales, we define three prompt-controlled reasoning modes. No-think directly emits the target SIDs. Low-think produces a concise analysis focusing on the decisive visual and textual constraints. High-think permits a more extensive reasoning trace for relational reasoning, constraint resolution, and candidate comparison. Mixed-budget supervised fine-tuning (SFT) jointly optimizes these reasoning behaviors. This stage enables the model to address diverse retrieval intents under different reasoning budgets, although the reasoning mode is still specified through the system prompt.

Mixing the intents and budgets at this stage gives RL a single
checkpoint in which alternative retrieval and reasoning behaviors are already
reachable.

\subsection{Intent-Balanced Unified RL}

Let $\mathcal{D}_t$ denote data for operational slice $t$. We sample from
\begin{equation}
    \mathcal{D}_{\mathrm{RL}}=\sum_{t=1}^{T}\beta_t\mathcal{D}_t,
    \qquad \sum_{t=1}^{T}\beta_t=1,
\end{equation}
where $\beta_t$ prevents high-volume intents from dominating optimization. For
a query from slice $t$, the calibrated evaluator applies the corresponding
reward rubric $r_t$. The slices select sampling and evaluation rules during
training, not experts or serving routes. Every rollout updates the same policy
$\pi_\theta$.

Under the prompt-specified reasoning budget, the policy generates a reasoning
trace and then the product SIDs. Group
relative policy optimization (GRPO) \citep{grpo} compares outcomes for the same
query. The following mechanisms address the training--serving
alignment.

\subsection{Sample--Beam Rollouts}
\label{sec:samplebeam}

Standard GRPO samples each completion autoregressively, whereas production
generative retrieval typically beam-decodes the SID portion. To reduce this
training--serving mismatch, we construct each rollout in two stages. Given a
query $x$ and reasoning-budget prompt $p$, we first sample $N$ reasoning
prefixes from the frozen old policy. For each prefix, we then run beam search
with width $B$ to generate $B$ SID sequences:
\begin{equation}
\begin{aligned}
h_g &\sim \pi_{\theta_{\mathrm{old}}}(\cdot\mid x,p),
    &&g=1,\ldots,N,\\
\{y_{g,b}\}_{b=1}^{B}
    &=\operatorname{Beam}_{B}\!\left(
      \pi_{\theta_{\mathrm{old}}}(\cdot\mid x,p,h_g)\right),\\
\tau_{g,b}&=h_g\oplus y_{g,b},
    &R_{g,b}&=r_t(x,y_{g,b}).
\end{aligned}
\label{eq:sample_beam}
\end{equation}
The group contains $N$ sampled reasoning prefixes and $NB$ beam-branch
trajectories. Each branch receives a candidate-level reward
$R_{g,b}=r_t(x,y_{g,b})$, and rewards are normalized across the $NB$
branches before applying a GRPO-style clipped surrogate to both reasoning and
SID tokens. The shared prefix $h_g$ therefore aggregates the advantages of its
$B$ SID branches. Beam selection is not differentiated. Since the SID branches
are selected by deterministic beam search rather than ancestral sampling, this
is a beam-conditioned surrogate rather than an unbiased on-policy estimator.
Its purpose is to align SID rollout generation, rather than the reward
function, with serving-time beam decoding.

\section{Scenario Scaling and Deployment}
\label{sec:scenario}

Intent describes what a user wants, scenario describes where the model runs and
under what resource envelope. We use one architecture and training framework to
construct native VLMs at $0.8$B, $2$B, and $4$B parameters. Model size provides
coarse-grained adaptation to an application's stable latency, memory, and
throughput constraints. Within a selected model, No-think, Low-think,
and High-think provide finer-grained operating points. Explicit budgets
are useful when predictable cost is required.

At serving time, the application passes the raw image and instruction to the
selected model. The model emits an optional reasoning trace, fixes the trace
when the \texttt{think} block closes, and beam-decodes a set of product SIDs.
Each valid SID maps deterministically to a catalog item, and the resulting
candidate set enters the application's retrieval or exposure interface. No
intent classifier selects an expert model. The application chooses model size,
reasoning policy, and beam configuration according to its resource and latency
envelope. These settings are held fixed when systems are compared online.

PailitaoMMSearch2.0 is deployed across multiple applications on Taobao Pailitao, including "\textit{Multimodal Search}", "\textit{Find Matching Tab}", and "\textit{Find Similarity Tab}". The applications share the request,
model, and SID interfaces while operating under different traffic and quality
requirements. Online interactions are sampled back into the real-traffic
workflow described in Section~\ref{sec:data}, where behavioral noise and query
corruption are removed before the records enter subsequent training. This
feedback path closes the data--training--deployment loop without changing the
serving-time requirement that the model infer the relation from the query.

The distinction between intent and scenario scaling avoids double-counting the
same mechanism.

\section{Experiments}
\label{sec:experiments}

\subsection{Setup}

\textbf{Data and splits.}
Training instances are produced by the workflows in
Section~\ref{sec:data} and organized into similar, editing, and matching
intents. The automatic testset is drawn from a later time window than the
training data, and duplicate query--target pairs are removed across splits.
The human-annotated evaluator-calibration set is disjoint from both training
and automatic evaluation. Table~\ref{tab:data} summarizes the resulting data
statistics.

\begin{table}[t]
\centering
\caption{Dataset statistics for training, automatic evaluation, and evaluator
calibration.}
\label{tab:data}
\vspace{-10pt}
\small
\begin{tabular}{lrrr}
\toprule
 & Similar & Editing & Matching\\
\midrule
Proxy/cold-start training & 24M & 24M & 12M\\
Refined real-traffic training & 1M & 8M & 1M\\
Automatic test & 502 & 772 & 371\\
Evaluator calibration & 40,000 & 20,000 & 40,000\\
\bottomrule
\end{tabular}
\vspace{-10pt}
\end{table}

\textbf{Metrics and evaluator calibration.}
For each query, a human-calibrated multimodal evaluator scores the top-10
results from 0 to 3. We report Good Rate (GR), the proportion scoring at least
2; Excellent Rate (ER), the proportion scoring 3; and NDCG@10 based on graded
relevance. All metrics are macro-averaged over the three intents. The evaluator
is calibrated on human annotations disjoint from both model-training and
automatic-test data, using the same intent-specific relevance criteria.
Table~\ref{tab:judge} shows that precision ranges from $92.4\%$ to $93.5\%$
for Good predictions and from $93.6\%$ to $93.9\%$ for Excellent predictions.
We use ER as the primary ablation metric and retain GR and NDCG@10 for the
overall comparison. Complementarily, the online A/B tests in
Section~\ref{sec:online} evaluate the deployed system through real user
interactions and business outcomes rather than evaluator scores.

\begin{table}[t]
\centering
\caption{Automatic-evaluator precision against disjoint human annotations.}
\label{tab:judge}
\vspace{-10pt}
\small
\begin{tabular}{lcc}
\toprule
Intent & Good Prec. ($\geq2$) & Excellent Prec. ($=3$)\\
\midrule
Similar & 0.924 & 0.936 \\
Editing & 0.935 & 0.939 \\
Matching & 0.932 & 0.936 \\
\bottomrule
\end{tabular}
\vspace{-12pt}
\end{table}

\textbf{Implementation.}
Unless otherwise stated, controlled experiments use the $2$B model with High-think mode. Variants
within each ablation share the same evaluation set, decoding constraints, and
training configuration except for the factor under study. \emph{Note that the choice of quality-control module and reward model is flexible, and we use the popular Qwen3.5-397B-A17 as quality-control, Qwen3.8-27B as reward model. The different base model configurations avoid the player-referee conflict.}

\subsection{Overall Multi-Intent Retrieval Quality}

We compare PailitaoMMSearch2.0 with two advanced VLM query-rewrite plus
full-text retrieval pipelines and an internal multimodal recall-and-ranking
pipeline. The internal baseline follows a multi-recall architecture. An
explicit intent classifier activates the appropriate channels, whose candidates
are merged from several SFT models by collected domain data, such as 1) composed image retrieval (CIR) by Qwen3-VL-Embedding-2B~\citep{li2026qwen3}, 2) multi-modal rewriting for text search by Qwen3-VL-2B~\citep{Qwen3-VL} and Taobao Text Search, 3)
and QueryText to ItemTitle semantic matching by DeepSeek~\citep{liu2024deepseek}. The merged candidate
set is processed by multiple ranking stages, ending with a large Transformer
ranker upgraded from the MTGR architecture~\citep{mtgr} that optimized for multimodal search.
Table~\ref{tab:main} reports intent-level and macro-averaged NDCG, GR, and ER,
together with average response time and GPU usage. The PailitaoMMSearch2.0 variants
use Low-think decoding to represent practical deployment configurations.

\begin{table*}[t]
\centering
\caption{Offline retrieval quality and serving cost across the three intents.
Higher NDCG, GR, and ER are better; lower average response time (RT) and fewer
GPU devices are better. Best and second-best quality and RT results are shown
in bold and underlined, respectively.}
\vspace{-10pt}
\label{tab:main}
\scriptsize
\setlength{\tabcolsep}{2pt}
\resizebox{\textwidth}{!}{%
\begin{tabular}{@{}l*{12}{c}rr@{}}
\toprule
\multirow{2}{*}{System}
& \multicolumn{3}{c}{Similar}
& \multicolumn{3}{c}{Editing}
& \multicolumn{3}{c}{Matching}
& \multicolumn{3}{c}{Macro}
& \multirow{2}{*}{\makecell{Avg. RT\\(ms)}}
& \multirow{2}{*}{\makecell{GPU\\devices}} \\
\cmidrule(lr){2-4}
\cmidrule(lr){5-7}
\cmidrule(lr){8-10}
\cmidrule(lr){11-13}
& NDCG & GR & ER
& NDCG & GR & ER
& NDCG & GR & ER
& NDCG & GR & ER
& & \\
\midrule
Qwen3.8-Flash-Next
& 0.885 & \textbf{74.5\%} & 47.5\%
& \underline{0.894} & \textbf{66.7\%} & \underline{47.9\%}
& 0.892 & 74.4\% & 61.1\%
& 0.890 & \textbf{71.9\%} & 52.2\%
& 1{,}075 & 4 \\
GPT6-luna
& 0.883 & \underline{71.3\%} & 44.7\%
& 0.859 & 60.0\% & 43.5\%
& 0.930 & 80.4\% & 67.3\%
& 0.891 & 70.6\% & 51.8\%
& 3{,}646 & -- \\
Internal recall-and-ranking pipeline
& 0.874 & 70.6\% & \textbf{54.1\%}
& \textbf{0.903} & \underline{64.0\%} & \textbf{50.1\%}
& 0.745 & 51.4\% & 44.2\%
& 0.841 & 62.0\% & 49.5\%
& 780 & -- \\
\midrule
PailitaoMMSearch2.0-0.8B $_{\textit{low-think}}$
& \underline{0.887} & 64.3\% & 44.6\%
& 0.877 & 51.2\% & 35.5\%
& \underline{0.944} & 81.1\% & 70.9\%
& 0.903 & 65.5\% & 50.3\%
& \textbf{187} & 1 \\
PailitaoMMSearch2.0-2B $_{\textit{low-think}}$
& \textbf{0.907} & 70.6\% & 50.5\%
& 0.891 & 57.4\% & 40.6\%
& \textbf{0.955} & \textbf{84.7\%} & \textbf{74.3\%}
& \textbf{0.918} & \underline{70.9\%} & \textbf{55.1\%}
& \underline{278} & 1 \\
PailitaoMMSearch2.0-4B $_{\textit{low-think}}$
& \textbf{0.907} & 69.5\% & \underline{50.6\%}
& 0.885 & 57.0\% & 40.6\%
& \underline{0.944} & \underline{83.6\%} & \underline{73.9\%}
& \underline{0.912} & 70.0\% & \underline{55.0\%}
& 617 & 1 \\
\bottomrule
\end{tabular}%
}
\vspace{-8pt}
\end{table*}

Among the PailitaoMMSearch2.0 variants, the $2$B Low-think configuration provides
the strongest aggregate operating point, achieving the highest macro NDCG
($0.918$) and macro ER ($55.1\%$) in the table with an average RT of $278$ ms
on one GPU. Relative to Qwen3.8-Flash-Next, GPT6-luna, and the internal
recall-and-ranking pipeline, it improves macro ER by $2.9$, $3.3$, and $5.6$
points while reducing average RT by $74.1\%$, $92.4\%$, and $64.4\%$,
respectively. It also uses one GPU, compared with four for
Qwen3.8-Flash-Next.

The largest advantage occurs on matching retrieval: the $2$B model reaches an
NDCG of $0.955$ and an ER of $74.3\%$, exceeding the matching ER of the two
rewrite-based systems by $13.2$ and $7.0$ points and that of the internal
pipeline by $30.1$ points. The gains are not uniform across intents. The
internal pipeline remains stronger on similar and editing ER, and both
rewrite-based systems also obtain higher editing ER than the $2$B model. 

Within
our model family, scaling from $0.8$B to $2$B raises macro ER by $4.8$ points
for only $91$ ms of additional RT. Increasing the model to $4$B does not
improve aggregate quality under Low-think: its macro ER is $0.1$ points lower
than the $2$B model while its RT increases from $278$ to $617$ ms. Although
the $4$B model is slightly weaker than the $2$B model under Low-think, it
benefits more from a larger reasoning budget and achieves higher macro ER under
High-think, as shown in Table~\ref{tab:res_budget}. These results identify the
$2$B Low-think configuration as the best overall quality--latency trade-off,
with particularly strong performance on matching-oriented requests.

\subsection{Effects of Data Scaling}

We first vary the accepted training corpus while preserving the distributions
of data sources and intents. Scaling the data volume from $30\%$ to $100\%$
raises macro ER from $46.32\%$ to $49.03\%$ ($+2.71$ points). Similar retrieval
benefits the most, improving by $5.08$ points, compared with gains of $1.70$
and $1.33$ points for editing and matching retrieval, respectively. The gains
are not uniform across scales or intents: macro ER increases by $1.49$ points
from $30\%$ to $70\%$ and by a further $1.22$ points at $100\%$, while matching
ER first decreases by $0.52$ points and then recovers by $1.85$ points. Thus,
more accepted data improves aggregate performance, but its marginal value is
intent dependent.

To separate data composition from volume, the final two rows of
Table~\ref{tab:data_ablation} ablate real-traffic supervision and quality
control at the full data scale. Removing real-traffic data lowers macro ER by
$0.78$ points, with losses of $1.10$, $0.86$, and $0.38$ points on similar,
editing, and matching retrieval, respectively. Further removing quality control
lowers macro ER by another $0.28$ points, driven mainly by matching ($0.94$
points) and editing ($0.30$ points), despite a $0.42$-point increase on similar
retrieval. Removing both components reduces macro ER by $1.06$ points and
degrades all three intents relative to the complete pipeline. Thus, real-traffic
supervision provides broad post-deployment gains, whereas quality control mainly
benefits aggregate and matching performance.

\begin{table}[t]
\centering
\caption{Effect of training-data scale and data-production components on the
$2$B SFT model.}
\vspace{-10pt}
\label{tab:data_ablation}
\small
\setlength{\tabcolsep}{3.5pt}
\begin{tabular}{@{}llcccc@{}}
\toprule
\makecell[l]{Data\\volume} &
\makecell[l]{Data\\setting} &
\makecell{Sim.\\ER} &
\makecell{Edit\\ER} &
\makecell{Match\\ER} &
\makecell{Macro\\ER} \\
\midrule
30\%  & -- & 40.48\% & 33.73\% & 64.76\%
             & 46.32\%\diff{-2.71\,pp} \\
70\%  & -- & 43.91\% & 35.29\% & 64.24\%
             & 47.81\%\diff{-1.22\,pp} \\
100\% & -- & 45.56\% & 35.43\% & 66.09\%
             & \textbf{49.03\%} \\
\midrule
100\% & w/o real traffic
             & 44.46\% & 34.57\% & 65.71\%
             & 48.25\%\diff{-0.78\,pp} \\
100\% & \makecell[l]{w/o real traffic\\w/o quality control}
             & 44.88\% & 34.27\% & 64.77\%
             & 47.97\%\diff{-1.06\,pp} \\
\bottomrule
\end{tabular}
\vspace{-10pt}
\end{table}

\subsection{Effects of Intent Scaling}

\textbf{Continual pre-training.}
We compare the Base VLM with a model that undergoes continual pre-training
(CPT) for e-commerce product understanding. Both initializations use the same
downstream SFT data and configuration. As shown in Table~\ref{tab:cpt}, CPT
improves macro GR by $0.8$ points and macro ER by $1.2$ points, showing that
domain adaptation before instruction tuning provides a stronger initialization
for the three retrieval intents.

\begin{table}[t]
\centering
\caption{Effect of continual pre-training under a fixed downstream SFT
configuration.}
\vspace{-10pt}
\label{tab:cpt}
\small
\begin{tabular}{lcc}
\toprule
Initialization & Macro GR & Macro ER \\
\midrule
Base VLM $+$ SFT
& 62.5\%\diff{-0.8\,pp}
& 47.8\%\diff{-1.2\,pp} \\
CPT $+$ SFT
& \textbf{63.3\%}
& \textbf{49.0\%} \\
\bottomrule
\end{tabular}
\vspace{-10pt}
\end{table}

\textbf{Cross-intent transfer.}
Table~\ref{tab:transfer} expands the SFT intent mixture in stages. Adding
similar-retrieval supervision to editing-only training raises editing ER from
$33.11\%$ to $35.88\%$ ($+2.77$ points), demonstrating positive transfer from
similar retrieval to editing. Introducing matching supervision then achieves
$66.09\%$ matching ER and further improves similar ER from $44.63\%$ to
$45.56\%$ ($+0.93$ points). Editing ER decreases slightly from $35.88\%$ to
$35.43\%$ ($-0.45$ points), but remains $2.32$ points above the editing-only
setting. The three-intent mixture therefore largely retains the established
capabilities while adding strong matching performance, although the transfer
is asymmetric rather than uniformly positive across intents.

\begin{table}[t]
\centering
\caption{Cross-intent transfer as the SFT supervision mixture expands.}
\vspace{-10pt}
\label{tab:transfer}
\small
\begin{tabular*}{\linewidth}{
    @{\extracolsep{\fill}}lccc@{}
}
\toprule
\makecell[l]{SFT intent\\mixture}
& \makecell{Similar\\ER}
& \makecell{Editing\\ER}
& \makecell{Matching\\ER} \\
\midrule
Editing only
& -- & 33.11\% & -- \\
Similar $+$ Editing
& 44.63\% & 35.88\% & -- \\
Similar $+$ Editing $+$ Matching
& \textbf{45.56\%} & \textbf{35.43\%} & \textbf{66.09\%} \\
\bottomrule
\end{tabular*}
\vspace{-10pt}
\end{table}

\textbf{Unified reinforcement learning.}
Table~\ref{tab:rl} separates the effect of standard unified RL from the
additional contribution of Sample--Beam rollouts. Standard RL raises macro ER
from $49.03\%$ to $52.70\%$ ($+3.67$ points), with gains of $1.23$, $1.94$, and
$7.84$ points on similar, editing, and matching retrieval, respectively.
Replacing standard rollouts with Sample--Beam further improves macro ER by
$2.30$ points, adding $2.86$ points on similar retrieval, $3.11$ points on
editing, and $0.95$ points on matching. Overall, Sample--Beam RL outperforms
SFT by $5.97$ macro-ER points, corresponding to intent-level gains of $4.09$,
$5.05$, and $8.79$ points on similar, editing, and matching retrieval,
respectively. The largest end-to-end gain occurs on matching, whereas the
incremental benefit of Sample--Beam over standard RL is concentrated on
editing and similar retrieval. This progression indicates that unified reward
optimization and beam-aligned rollouts provide complementary improvements,
although the incremental effect of Sample--Beam varies across
intents.

\begin{table}[t]
\centering
\caption{Effects of unified RL and Sample--Beam rollouts under matched
initialization, intent mixture, and reward evaluation.}
\label{tab:rl}
\small
\setlength{\tabcolsep}{3pt}
\begin{tabular*}{\linewidth}{
    @{\extracolsep{\fill}}lcccc@{}
}
\toprule
\makecell[l]{Training\\setting}
& \makecell{Similar\\ER}
& \makecell{Editing\\ER}
& \makecell{Matching\\ER}
& \makecell{Macro\\ER} \\
\midrule
SFT
& 45.56\%
& 35.43\%
& 66.09\%
& 49.03\%\diff{-5.97\,pp} \\
Standard RL
& 46.79\%
& 37.37\%
& 73.93\%
& 52.70\%\diff{-2.3\,pp} \\
Sample--Beam RL
& \textbf{49.65\%}
& \textbf{40.48\%}
& \textbf{74.88\%}
& \textbf{55.00\%} \\
\bottomrule
\end{tabular*}
\end{table}

\subsection{Effects of Scenario Scaling}

\begin{figure}[t]
\centering
\includegraphics[width=\columnwidth]{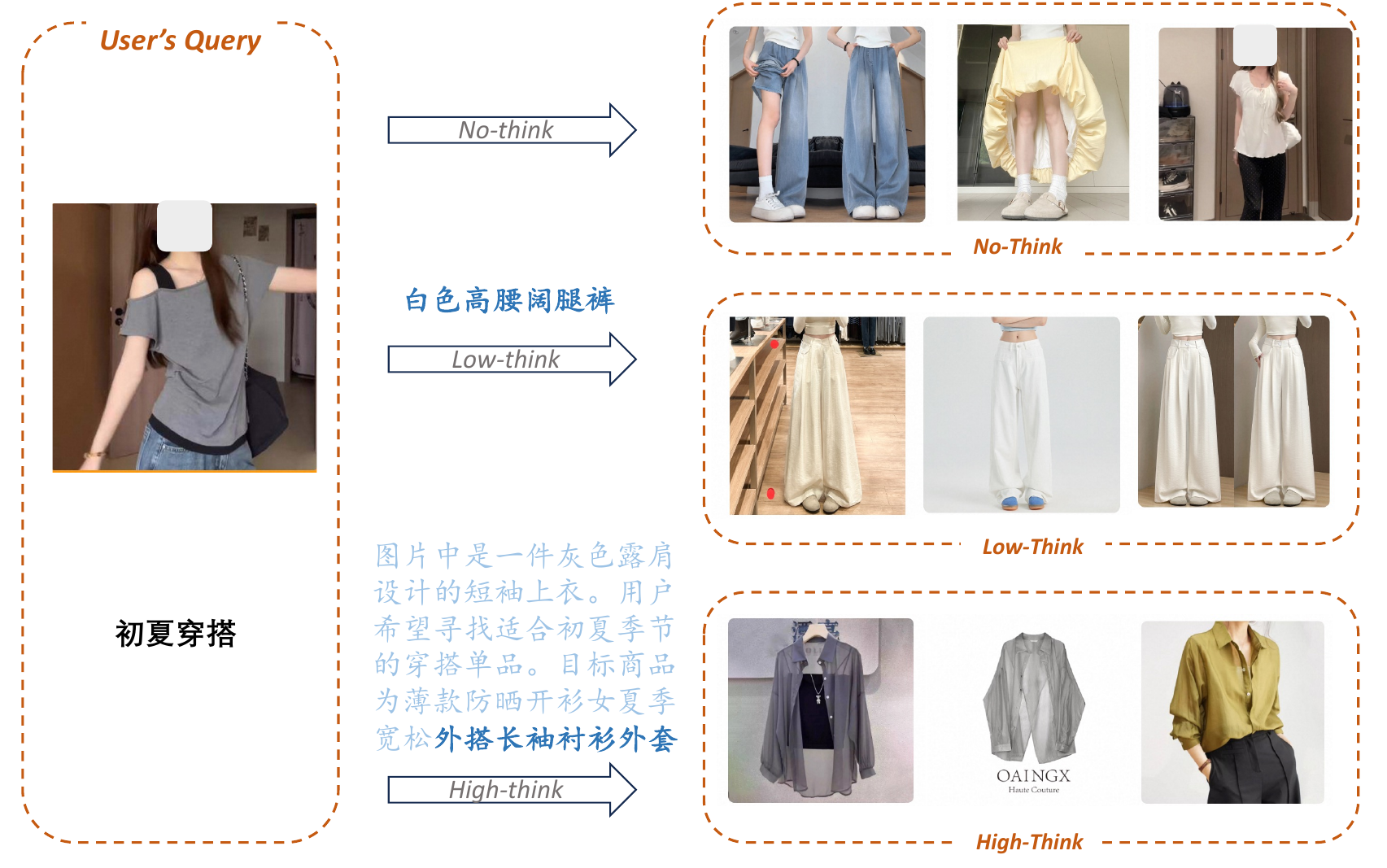}
\vspace{-12pt}
\caption{Case study of the three reasoning budgets on representative queries.
No-think answers directly, Low-think isolates the decisive visual and textual
constraints, and High-think adds a longer trace. The examples illustrate the effect of the extra reasoning.}
\label{fig:think_case}
\vspace{-12pt}
\end{figure}


Table~\ref{tab:res_budget} compares retrieval quality, reasoning tokens, and
response time across three model scales and inference-time budgets. For the
$0.8$B model, No-think achieves the highest macro ER ($50.78\%$) at the lowest
latency ($136$ ms); additional reasoning reduces macro ER to $50.31\%$ with
Low-think and $49.60\%$ with High-think while increasing serving cost. For the
$2$B model, Low-think provides the best aggregate trade-off, improving macro ER
from $54.62\%$ to $55.13\%$ with only $8.6$ reasoning tokens and $278$ ms RT.
High-think further raises matching ER to $74.88\%$, but its macro ER decreases
by $0.13$ points relative to Low-think and its RT nearly doubles to $544$ ms.
The $4$B model benefits most from additional reasoning: Low-think improves
macro ER by $0.98$ points over No-think, while High-think reaches the best
macro ER ($55.41\%$), editing ER ($41.27\%$), and matching ER ($75.57\%$) in
the table, at the cost of $54.99$ reasoning tokens and $1{,}133$ ms RT.
The appropriate
operating point therefore depends on the target intent and quality--latency
requirements rather than always selecting the largest model or budget.
Figure~\ref{fig:think_case} shows representative queries under each budget and
illustrates where additional reasoning changes the retrieved set.

\begin{table}[t]
\centering
\caption{Effects of model scale and inference-time reasoning budget on
retrieval quality, reasoning tokens, and response time. Quality columns report
ER.}
\label{tab:res_budget}
\scriptsize
\setlength{\tabcolsep}{2pt}
\resizebox{\columnwidth}{!}{%
\begin{tabular}{@{}llrrrrrr@{}}
\toprule
Params & Budget
& \makecell{Similar\\ER}
& \makecell{Editing\\ER}
& \makecell{Matching\\ER}
& \makecell{Macro\\ER}
& \makecell{Avg.\\tokens}
& \makecell{RT\\(ms)} \\
\midrule
\multirow{3}{*}{$0.8$B}
& No-think   & 46.22\% & 35.67\% & 70.45\% & 50.78\% & 0     & 136 \\
& Low-think  & 44.55\% & 35.51\% & 70.87\% & 50.31\% & 9.0   & 187 \\
& High-think & 43.51\% & 35.09\% & 70.21\% & 49.60\% & 51.7  & 326 \\
\midrule
\multirow{3}{*}{$2$B}
& No-think   & 49.71\% & \underline{40.70\%} & 73.45\% & 54.62\% & 0     & 210 \\
& Low-think  & \underline{50.54\%} & 40.60\% & 74.25\% & \underline{55.13\%} & 8.6   & 278 \\
& High-think & 49.65\% & 40.48\% & \underline{74.88\%} & 55.00\% & 54.4  & 544 \\
\midrule
\multirow{3}{*}{$4$B}
& No-think   & 50.11\% & 40.20\% & 71.83\% & 54.05\% & 0     & 465 \\
& Low-think  & \textbf{50.62\%} & 40.58\% & 73.90\% & 55.03\% & 8.9   & 617 \\
& High-think & 49.39\% & \textbf{41.27\%} & \textbf{75.57\%} & \textbf{55.41\%} & 54.99 & 1{,}133 \\
\bottomrule
\end{tabular}%
}
\vspace{-10pt}
\end{table}

\subsection{Online A/B Tests in Production}
\label{sec:online}

We deploy PailitaoMMSearch2.0 in Pailitao applications and conduct
controlled online A/B tests against their respective production baselines. For
each experiment, approximately $10\%$ of the online traffic is allocated to
the experimental bucket, and each test runs for approximately 14 days. In
Pailitao ``\textit{Multimodal Search}'', PailitaoMMSearch2.0 increases item page
views by $\textbf{17.90\%}$, unique-user item page views by $\textbf{13.62\%}$, transactions by
$\textbf{27.04\%}$, gross merchandise value (GMV) by $\textbf{8.94\%}$, click-through rate by
$\textbf{0.58}$ percentage points, and conversion rate by $\textbf{0.37}$ percentage points.
All improvements are statistically significant ($p < 0.05$).

Across these applications, PailitaoMMSearch2.0 operates as an end-to-end retrieval
component: it consumes the user's raw image--text query and directly produces
the final product list. For applications with an existing mature
recall--ranking pipeline, products returned by PailitaoMMSearch2.0 are interleaved
with those from the existing pipeline at the final presentation stage. The
online measurements therefore evaluate the complete deployed configuration,
whereas the controlled offline ablations provide component-level analysis.
Following the statistically significant gains observed in the online
experiments, PailitaoMMSearch2.0 has been fully rolled out to $100\%$ of the
traffic in the corresponding production scenarios.

\section{Conclusion, Limitations, and Future Work}
\label{sec:conclusion}

We presented PailitaoMMSearch2.0, a unified end-to-end multimodal generative
retrieval system for free-form image--text queries and multiple shopping
relations. Beyond a retrieval model, PailitaoMMSearch2.0 provides a practical,
industrial-scale blueprint for open-ended multimodal search, spanning multimodal
training-data synthesis and quality control, multi-stage model training, and
production serving. A composable data-production toolset supports cold start and
real-traffic refinement, while CPT, mixed-intent SFT, and unified RL train one
model across similar, editing, and matching retrieval. Offline experiments show
benefits from data scaling, domain-adaptive CPT, cross-intent supervision, and
Sample--Beam RL. The reasoning-budget study further identifies a practical
quality--latency frontier, and online A/B tests demonstrate system-level gains
in production.

Our study has several limitations. The term ``open-ended'' refers to the query
interface rather than universal relation coverage, and our evaluation spans only
three shopping relations in one e-commerce domain. Most ablations use the $2$B
checkpoint, and the three model sizes are separate checkpoints rather than one
elastic set of weights, so their interactions remain underexplored. Offline
results also rely on an automatic judge that, despite calibration against
disjoint human labels, may share correlated errors with the evaluators used for
data filtering and reward modeling. Finally, the online A/B test measures the
deployed system as a whole and cannot attribute gains to individual components.

Future work will broaden the relation taxonomy, jointly optimize retrieval
quality against latency and token cost, and study adaptive or distilled models
that share parameters across serving budgets. We also plan to pair evaluator
ensembles with periodic human audits and to examine stability under catalog and
traffic drift.

\section{Acknowledgement}
We would like to express our gratitude to the following teams for their invaluable contributions to this work:
\begin{itemize}
    \item The \textbf{AI Infra} team, for inference acceleration and optimization during model deployment.
    \item The \textbf{Engineering Development} team and \textbf{Product} team, for their support in usage and product design throughout the online deployment process.
    \item The other \textbf{Team Members}, for their support and cooperation in daily work.
\end{itemize}

\bibliographystyle{ACM-Reference-Format}
\bibliography{references}

\newpage

\appendix

\twocolumn[
\begin{minipage}{\textwidth}
\centering
\captionof{table}{Evolution of the editing-intent evaluation rubric. Each
revision was motivated by recurring disagreements with held-out human
annotations.}
\label{tab:rubric-evolution}
\small
\setlength{\tabcolsep}{4pt}
\begin{tabular}{p{0.15\textwidth}p{0.22\textwidth}p{0.27\textwidth}p{0.28\textwidth}}
\toprule
Stage & Evaluation focus & Recurring failure & Revision introduced \\
\midrule
Broad relevance
& Judge whether the target is generally related to the query image and text.
& A visually similar item can receive a high score even when it does not perform
the requested edit.
& Require the evaluator to identify the intended target category and requested
transformation before assigning a score. \\
\addlinespace
Relation-aware
& Verify the requested edit while preserving explicitly constrained attributes.
& The main edit can be correct even though the target directly violates another
explicit requirement.
& Separate requested changes and preservation requirements from attributes not
constrained by the instruction. \\
\addlinespace
Calibrated final
& Jointly assess fine-grained category correctness, explicit requirements, and
consistency with the editing direction.
& Borderline cases depend on whether the evidence is insufficient or an explicit
requirement is directly contradicted.
& Adopt the four-level decision rules in Table~\ref{tab:editing-score-rubric} and
calibrate them with intent-specific examples. \\
\bottomrule
\end{tabular}

\vspace{4pt}
\captionof{table}{Final four-level relevance rubric for the editing intent.}
\label{tab:editing-score-rubric}
\setlength{\tabcolsep}{4pt}
\begin{tabular}{cp{0.18\textwidth}p{0.68\textwidth}}
\toprule
Score & Interpretation & Decision rule \\
\midrule
3 & Fully consistent
& The target belongs to the correct fine-grained product category, and clear
evidence supports every explicit requirement. Its changes relative to the query
product closely match the editing direction expressed by the instruction. \\
\addlinespace
2 & Mostly consistent
& The target belongs to the correct fine-grained category and does not directly
violate any explicit requirement. The core edit is satisfied, but some evidence
is inconclusive, or unconstrained style differences make the overall editing
consistency weaker than that of a score-3 target. \\
\addlinespace
1 & Explicitly inconsistent
& The target belongs to the correct fine-grained category, but clear evidence in
its image or title directly violates at least one explicit editing or
preservation requirement. A score of 1 must not be assigned unless the violated
requirement can be identified in the instruction. \\
\addlinespace
0 & Invalid or category-incorrect
& The text does not express a valid product-retrieval intent, or the target does
not belong to the intended fine-grained category and differs in product form or
primary function. If the instruction explicitly requests a new target category,
the target is evaluated against that category rather than the category of the
original product. \\
\bottomrule
\end{tabular}
\end{minipage}
\vspace{6pt}
]

\section{Evolution of the Editing-Intent Rubric}
\label{app:rubric-evolution}

This appendix documents how we refined the automatic relevance rubric for
editing-oriented retrieval. An editing query specifies a controlled
transformation of the query product---for example, changing sleeve length while
preserving its color. The evaluator must determine whether the target belongs to
the intended fine-grained product category, satisfies the explicit editing and
preservation requirements, and remains consistent with the requested editing
direction.

\subsection{Rubric Refinement}

Early rubric versions emphasized visual and semantic similarity. Error analysis
against held-out human annotations showed that this criterion was insufficient:
a visually similar product could score highly without performing the requested
edit, while a target that performed the main edit could still directly violate
an explicit preservation requirement. Table~\ref{tab:rubric-evolution}
summarizes the resulting revisions.

\subsection{Final Scoring Rule}

The final rubric first identifies the intended fine-grained target category and
extracts all explicit editing and preservation requirements from the
instruction. It then evaluates the target using evidence from both its image and
title and assigns one of the four scores in
Table~\ref{tab:editing-score-rubric}.

For example, given an image of a red short-sleeve shirt and the instruction
``find a long-sleeve one in the same color,'' a clearly red long-sleeve shirt in
the intended fine-grained category receives a score of 3. A red shirt that
appears to have long sleeves but for which the sleeve evidence is partially
obscured may receive a score of 2, provided that no explicit requirement is
directly contradicted. A clearly blue long-sleeve shirt receives a score of 1
because it directly violates the explicit color-preservation requirement. A
long-sleeve jacket or an unrelated product receives a score of 0 because it
does not belong to the intended fine-grained product category.

During data production, targets with scores of 2 or 3 pass the relevance gate,
score 3 is additionally used to compute the Excellent Rate reported in the
experiments. The tables present the decision structure rather than the complete
production prompt, which also includes intent-specific demonstrations and
output-format instructions.

\end{document}